\documentclass[a4paper]{article}

\usepackage{INTERSPEECH2021}

\usepackage{times}
\usepackage{latexsym}
\usepackage{multirow}

\usepackage{times}
\usepackage{latexsym}
\usepackage{multirow}

\usepackage{graphicx}
\usepackage{multirow, boldline}
\usepackage{subfigure}

\usepackage{soulutf8}
\usepackage{todonotes}
\usepackage{placeins}
\usepackage{tikz}
\usepackage{svg}
\usepackage{hyperref}

\usepackage{mathtools}

\makeatletter
\newcommand*\concat
  {\mathbin{\mathmakebox[\widthof{${+}\m@th$}]{+\hskip-1emplus1fil+}}}

\makeatother

 \newcommand\citep{\cite}

\title{YourTTS: Towards Zero-Shot Multi-Speaker TTS and Zero-Shot Voice Conversion for everyone}
\name{Edresson Casanova$^1$, Julian Weber$^2$, Christopher Shulby$^3$,  Arnaldo Candido Junior$^4$, \\ Eren Gölge$^5$ and  Moacir Antonelli Ponti$^1$}

\address{
    $^1$ Instituto de Ciências Matemáticas e de Computação, Universidade de S\~ao Paulo, Brazil \\
    $^2$ Sopra Banking Software, France \\
    $^3$ Defined.ai, United States of America \\
    $^4$ Federal University of Technology – Paraná, Brazil  \\
    $^5$ Coqui, Germany
  }
  
\email{edresson@usp.br}

\begin{document}
\maketitle

\begin{abstract}
   YourTTS brings the power of a multilingual approach to the task of zero-shot multi-speaker TTS. Our method builds upon the VITS model and adds several novel modifications for zero-shot multi-speaker and multilingual training. We achieved state-of-the-art (SOTA) results in zero-shot multi-speaker TTS and results comparable to SOTA in zero-shot voice conversion on the VCTK dataset. Additionally, our approach achieves promising results in a target language with a single-speaker dataset, opening possibilities for zero-shot multi-speaker TTS and zero-shot voice conversion systems in low-resource languages. Finally, it is possible to fine-tune the YourTTS model with less than 1 minute of speech and achieve state-of-the-art results in voice similarity and with reasonable quality. This is important to allow synthesis for speakers with a very different voice or recording characteristics from those seen during training.
\end{abstract}


\section{Introduction}
\label{sec:intro}
Text-to-Speech (TTS) systems have significantly advanced in recent years with deep learning approaches, allowing successful applications such as speech-based virtual assistants.  Most TTS systems were tailored from a single speaker's voice, but there is current interest in synthesizing voices for new speakers (not seen during training), employing only a few seconds of speech. This approach is called zero-shot multi-speaker TTS (ZS-TTS) as in \cite{jia2018transfer, cooper2020zero, choi2020attentron, casanova2021sc}.

ZS-TTS using deep learning was first proposed by 
\cite{arik2018neural} which extended the DeepVoice~3 method~\cite{deepvoice3}. Meanwhile, Tacotron 2~\cite{tacotron2} was adapted using external speaker embeddings extracted from a trained speaker encoder using a generalized end-to-end loss (GE2E)~\cite{ge2e}, allowing for speech generation that resembles the target speaker~\cite{jia2018transfer}. Similarly, Tacotron 2 was used with a different speaker embeddings methods~\cite{cooper2020zero}, with LDE embeddings~\cite{cai2018exploring} to improve similarity and naturalness of speech for unseen speakers~\cite{snyder2018x}. The authors also showed that a gender-dependent model improves the similarity for unseen speakers~\cite{cooper2020zero}. 
In this context, Attentron~\cite{choi2020attentron} proposed a fine-grained encoder with an attention mechanism for extracting detailed styles from various reference samples and a coarse-grained encoder. As a result of using several reference samples, they achieved better voice similarity for unseen speakers. { ZSM-SS \cite{kumar21c_interspeech} is a Transformer-based architecture with a normalization architecture and an external speaker encoder based on Wav2vec 2.0 \cite{baevski2020wav2vec}. The authors conditioned the normalization architecture with speaker embeddings, pitch, and energy. 
Despite promising results, the authors did not compare the proposed model with any of the related works mentioned above.} SC-GlowTTS~\cite{casanova2021sc} was the first application of flow-based models in ZS-TTS. It improved voice similarity for unseen speakers in training with respect to previous studies while maintaining  comparable quality. 

Despite these advances, the similarity gap between observed and unobserved speakers during training is still an open research question. ZS-TTS models still require a considerable amount of speakers for training, making it difficult to obtain high-quality models in low-resource languages. Furthermore, according to 
\cite{tan2021survey}, the quality of current ZS-TTS models is not sufficiently good, especially for target speakers with speech characteristics that differ from those seen in training. Although SC-GlowTTS~\cite{casanova2021sc} achieved promising results with only 11 speakers from the VCTK dataset \cite{veaux2016superseded}, when one limits the number and variety of training speakers, it also further hinders the model generalization for unseen voices.

In parallel with the ZS-TTS, multilingual TTS has also evolved aiming at learning models for multiple languages at the same time~ \cite{cao2019end,zhang2019learning,nekvinda2020one,li2021light}. Some of these models are particularly interesting as they allow for code-switching, i.e. changing the target language for some part of a sentence, while keeping the same voice \cite{nekvinda2020one}. This can be useful in ZS-TTS as it allows using of speakers from one language to be synthesized in another language.

In this paper, we propose YourTTS with several novel ideas focused on zero-shot multi-speaker and multilingual training. 
We report state-of-the-art zero-shot multi-speaker TTS results, as well as results comparable to SOTA in zero-shot voice conversion for the VCTK dataset.

Our novel zero-shot multi-speaker TTS approach includes the following contributions:

\begin{itemize}
    \item State-of-the-art results in the English Language;
    \item The first work proposing a multilingual approach in the zero-shot multi-speaker TTS scope;
    \item Ability to do zero-shot multi-speaker TTS and zero-shot Voice Conversion with promising quality and similarity in a target language using only one speaker in the target language during model training; 
     
     \item Require less than 1 minute of speech to fine-tune the model for speakers who have voice/recording characteristics very different from those seen in model training, and still achieve good similarity and quality results.
\end{itemize}

The audio samples for each of our experiments are available on the demo web-site\footnote{https://edresson.github.io/YourTTS/}. 
For reproducibility, our source-code is available at the Coqui TTS\footnote{https://github.com/coqui-ai/TTS}, as well as the model checkpoints of all experiments\footnote{https://github.com/Edresson/YourTTS}.

\section{YourTTS Model}\label{sec:TTSModel}

YourTTS builds upon VITS \cite{kim2021conditional}, but includes several novel modifications for zero-shot multi-speaker and multilingual training. First, unlike previous work \cite{casanova2021sc, kim2021conditional}, in our model we used raw text as input instead of phonemes. This allows more realistic results for languages without good open-source grapheme-to-phoneme converters available.

As in previous works, e.g.~\cite{kim2021conditional}, we use a transformer-based text encoder \cite{kim2020glow, casanova2021sc}.  However, for multilingual training, we concatenate 4-dimensional trainable language embeddings into the embeddings of each input character. In addition, we also increased the number of transformer blocks to 10 and the number of hidden channels to 196. As a decoder, we use a stack of 4 affine coupling layers \cite{dinh2016density} each layer is itself a stack of 4 WaveNet residual blocks \cite{oord2016wavenet}, as in VITS model.

As a vocoder we use the HiFi-GAN \cite{kong2020hifi} version 1 with the discriminator modifications introduced by \cite{kim2021conditional}. Furthermore, for efficient end2end training, we connect the TTS model with the vocoder using a variational autoencoder (VAE) \cite{kingma2013auto}. For this, we use the Posterior Encoder proposed by \cite{kim2021conditional}. The Posterior Encoder consists of 16 non-causal WaveNet residual blocks \cite{prenger2019waveglow, kim2020glow}. As input, the Posterior Encoder receives a linear spectrogram and predicts a latent variable, this latent variable is used as input for the vocoder and for the flow-based decoder, thus, no intermediate representation (such as mel-spectrograms) is necessary. This allows the model to learn an intermediate representation; hence, it achieves superior results to a two-stage approach system in which the vocoder and the TTS model are trained separately \cite{kim2021conditional}. Furthermore, to enable our model to synthesize speech with diverse rhythms from the input text, we use the stochastic duration predictor proposed in  \cite{kim2021conditional}. 

\begin{figure*}[t]
\centering
\resizebox{0.85\textwidth}{!}{%
\hfill
  \scriptsize
  \centering
\subfigure[Training procedure]{ \includegraphics[width=0.52\textwidth]{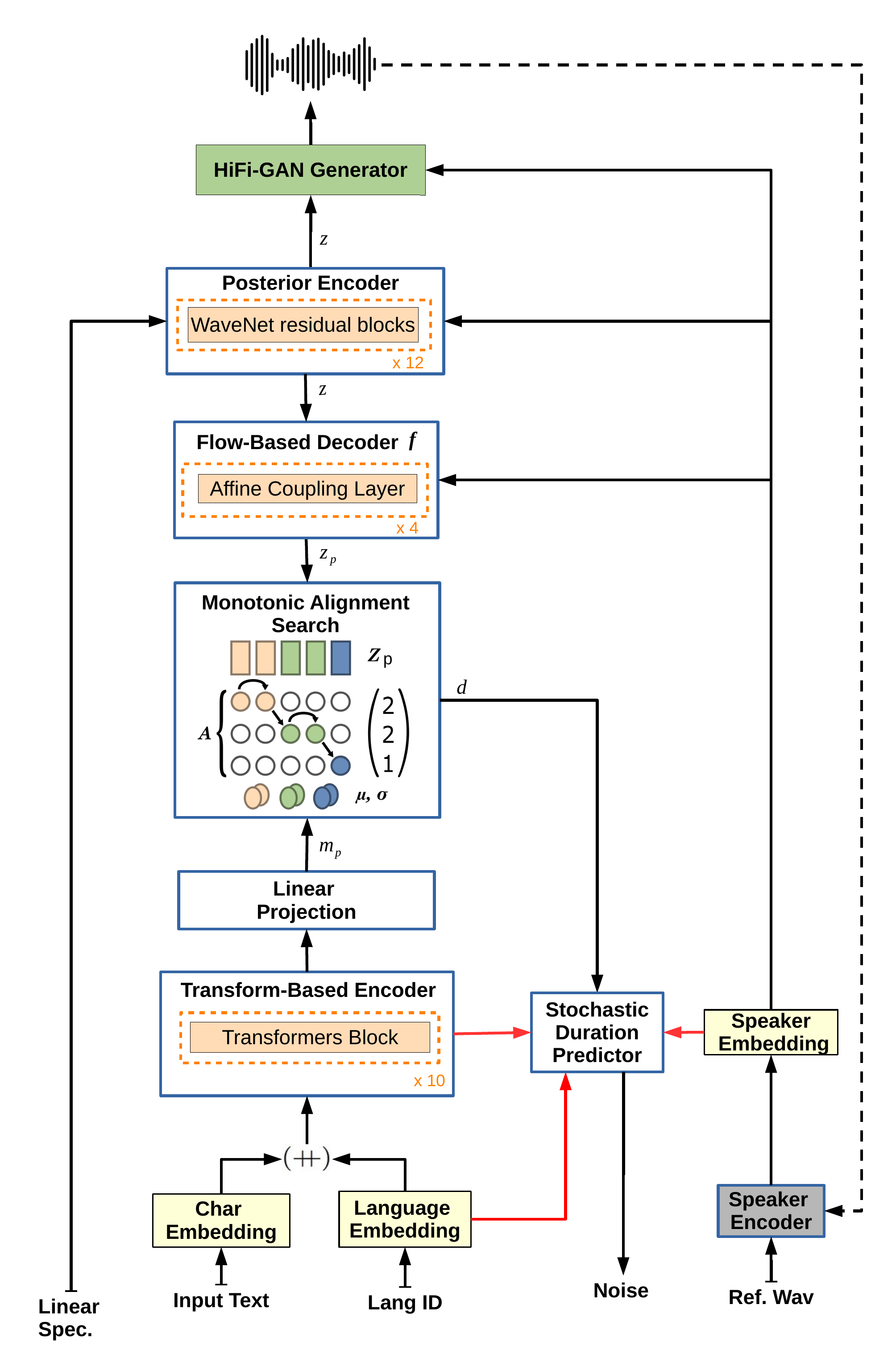}} 
\hfill
  \scriptsize
  \centering
\subfigure[Inference procedure]{ \includegraphics[width=0.49\textwidth]{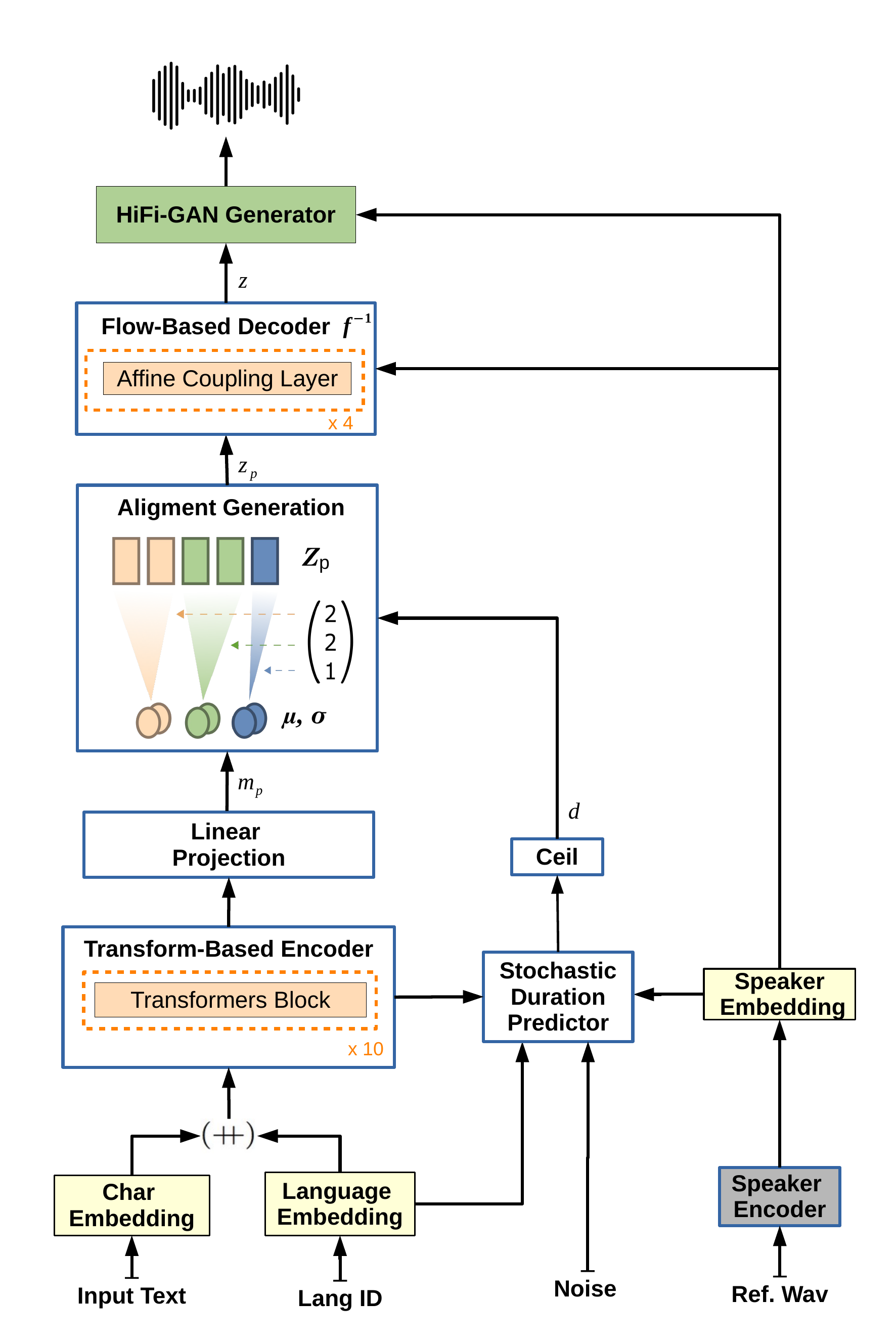}}
\hfill
}
 \caption{YourTTS diagram depicting (a) training procedure and (b) inference procedure.}
 \label{fig:generaltop}
\end{figure*}

YourTTS during training and inference is illustrated in Figure~\ref{fig:generaltop}, where $(\concat)$ indicates concatenation, red connections mean no gradient will be propagated by this connection, and dashed connections are optional. We omit the Hifi-GAN discriminator networks for simplicity.

To give the model zero-shot multi-speaker generation capabilities we condition all affine coupling layers of the flow-based decoder, the posterior encoder, and the vocoder on external speaker embeddings. We use global conditioning \cite{oord2016wavenet} in the residual blocks of the coupling layers as well as in the posterior encoder. We also sum the external speaker embeddings with the text encoder output and the decoder output before we pass them to the duration predictor and the vocoder, respectively. We use linear projection layers to match the dimensions before element-wise summations (see Figure \ref{fig:generaltop}). 



Also, inspired by \cite{xin21_interspeech}, we investigated Speaker Consistency Loss (SCL) in the final loss. In this case, a pre-trained speaker encoder is used to extract speaker embeddings from the generated audio and ground truth on which we maximize the cosine similarity. Formally, let $\phi(.)$ be a function outputting the embedding of a speaker, $cos\_sim$ be the cosine similarity function, $\alpha$ a positive real number that controls the influence of the SCL in the final loss, and $n$ the batch size, the SCL is defined as follows:
\begin{equation}
\label{eq:SCL}
     L_{SCL} =  \frac{- \alpha }{n} \cdot \sum_{i}^{n}{~cos\_sim(\phi(g_{i}), \phi(h_{i}))},
\end{equation}
where $g$ and $h$ represent, respectively, the ground truth and the generated speaker audio. 

During training, the Posterior Encoder receives linear spectrograms and speaker embeddings as input and predicts a latent variable $z$. This latent variable and speaker embeddings are used as input to the GAN-based vocoder generator which generates the waveform. For efficient end-to-end vocoder training, we randomly sample constant length partial sequences from $z$ as in~\cite{kong2020hifi, binkowski2019high, ren2020fastspeech, kim2021conditional}. The Flow-based decoder aims to condition the latent variable $z$ and speaker embeddings with respect to a $P_{Zp}$ prior distribution. To align the $P_{Zp}$ distribution with the output of the text encoder, we use the Monotonic Alignment Search (MAS)~\cite{kim2020glow, kim2021conditional}. The stochastic duration predictor receives as input speaker embeddings, language embeddings and the duration obtained through MAS. To generate human-like rhythms of speech, the objective of the stochastic duration predictor is a variational lower bound of the log-likelihood of the phoneme (pseudo-phoneme in our case) duration.

During inference, MAS is not used. Instead, $P_{Zp}$ distribution is predicted by the text encoder and the duration is sampled from random noise through the inverse transformation of the stochastic duration predictor and then, converted to integer. In this way, a latent variable $z_p$ is sampled from the distribution $P_{Zp}$. The inverted Flow-based decoder receives as input the latent variable $z_p$ and the speaker embeddings, transforming the latent variable $z_p$ into the latent variable $z$ which is passed as input to the vocoder generator, thus obtaining the synthesized waveform.

\section{Experiments} \label{sec:method}


\subsection{Speaker Encoder} \label{sec:SE}


%

As speaker encoder, we use the H/ASP model \cite{heo2020clova} publicly available, that was trained with the Prototypical Angular \cite{chung2020in} plus Softmax loss functions in the VoxCeleb 2 \cite{chung2018voxceleb2} dataset. This model was chosen for achieving state-of-the-art results in VoxCeleb 1 \cite{nagrani2017voxceleb} test subset. In addition, we evaluated the model in the test subset of Multilingual LibriSpeech (MLS) \cite{PratapXSSC20} using all languages. This model reached an average Equal Error Rate (EER) of 1.967 while the speaker encoder used in the SC-GlowTTS paper \cite{casanova2021sc} reached an EER of 5.244.

\subsection{Audio datasets}\label{sec:method:base}


We investigated 3 languages, using one dataset per language to train the model. For all datasets, pre-processing was carried out in order to have samples of similar loudness and to remove long periods of silence. All the audios to 16Khz and applied voice activity detection (VAD) using Webrtcvad toolkit\footnote{https://github.com/wiseman/py-webrtcvad} to trim the trailing silences. Additionally, we normalized all audio to -27dB using the RMS-based normalization from the Python package ffmpeg-normalize\footnote{https://github.com/slhck/ffmpeg-normalize}.

\textbf{English}: VCTK \cite{veaux2016superseded} dataset, which contains 44 hours of speech and 109 speakers, sampled at 48KHz. 
We divided the VCTK dataset into: train, development (containing the same speakers as the train set) and test. For the test set, we selected 11 speakers that are neither in the development nor the training set; following the proposal by \cite{jia2018transfer} and \cite{casanova2021sc}, we selected 1 representative from each accent totaling 7 women and 4 men (speakers 225, 234, 238, 245, 248, 261, 294, 302, 326, 335 and 347). 
Furthermore, in some experiments we used the subsets \textit{train-clean-100} and \textit{train-clean-360} of the LibriTTS dataset \cite{zen2019libritts} seeking to increase the number of speakers in the training of the models.

\textbf{Portuguese}: TTS-Portuguese Corpus \cite{casanova2020ttsportuguese}, a single-speaker dataset of the Brazilian Portuguese language with around 10 hours of speech, sampled at 48KHz. As the authors did not use a studio, the dataset contains ambient noise. We used the FullSubNet model \cite{Hao_2021} as denoiser and resampled the data to 16KHz. For development we randomly selected 500 samples and the rest of the dataset was used for training.

\textbf{French}: fr\_FR set of the M-AILABS dataset \cite{mailabs}, which is based on LibriVox\footnote{https://librivox.org/}. It consists of 2 female (104h) and 3 male speakers (71h) sampled at 16KHz.


To evaluate the zero-shot multi-speaker capabilities of our model in English, we use the 11 VCTK speakers reserved for testing. To further test its performance outside of the VCTK domain, we select 10 speakers (5F/5M) from subset \textit{test-clean} of LibriTTS dataset \cite{zen2019libritts}. For Portuguese we select samples from 10 speakers (5F/5M) from the Multilingual LibriSpeech (MLS) \cite{PratapXSSC20} dataset. For French, no evaluation dataset was used, due to the reasons described in Section \ref{sec:results}. Finally, for speaker adaptation experiments, to mimic a more realistic setting, we used 4 speakers from the Common Voice dataset \cite{ardila2020common}.


\subsection{Experimental setup}\label{sec:method:experiments}

We carried out four training experiments with YourTTS: 
\begin{itemize}
    \item \textbf{Experiment 1:}  using VCTK dataset (monolingual);
     \item \textbf{Experiment 2:} using both VCTK and TTS-Portuguese datasets (bilingual);
    \item \textbf{Experiment 3:} using VCTK, TTS-Portuguese and M-AILABS french datasets (trilingual);
    \item \textbf{Experiment 4:} 
    starting with the model obtained in experiment 3 we continue training with 1151 additional English speakers from both LibriTTS partitions \textit{train-clean-100} and \textit{train-clean-360}.
\end{itemize}

To accelerate training, in every experiment, we use transfer learning. In experiment 1, we start from a model trained 1M steps on LJSpeech \cite{ito2017lj} and continue the training for 200K steps with the VCTK dataset. However, due to the proposed changes, some layers of the model were randomly initialized due to the incompatibility of the shape of the weights. For experiments 2 and 3, training is done by continuing from the previous experiment for approximately 140k steps, learning one language at a time. In addition, for each of the experiments a fine-tuning was performed for 50k steps using the Speaker Consistency Loss (SCL), described in section \ref{sec:TTSModel}, with $\alpha = 9$. Finally, for experiment 4, we continue training from the model from experiment 3 fine-tuned with the Speaker Consistency Loss. 
Note that, although the latest works in ZS-TTS \cite{cooper2020zero, choi2020attentron, casanova2021sc} only use the VCTK dataset, this dataset has a limited number of speakers (109) and little variety of recording conditions. Thus, after training with VCTK only, in general, ZS-TTS models do not generalize satisfactorily to new speakers where recording conditions or voice characteristics are very different than those seen in the training \cite{tan2021survey}.

The models were trained using an NVIDIA TESLA V100 32GB with a batch size of 64. For the TTS model training and for the discrimination of vocoder HiFi-GAN we use the AdamW optimizer \cite{loshchilov2017decoupled}  with betas 0.8 and 0.99, weight decay 0.01 and an initial learning rate of 0.0002 decaying exponentially by a gamma of 0.999875~\cite{paszke2017automatic}. For the  multilingual experiments, we use weighted random sampling  \cite{paszke2017automatic} to guarantee a language balanced batch.

\section{Results and  Discussion} \label{sec:results}


In this paper, we evaluate synthesized speech quality using a Mean Opinion Score (MOS) study, as in \cite{mos}. To compare the similarity between the synthesized voice and the original speaker, we calculate the Speaker Encoder Cosine Similarity (SECS) \cite{casanova2021sc} between the speaker embeddings of two audios extracted from the speaker encoder. It ranges from -1 to 1, and a larger value indicates a stronger similarity \cite{cooper2020zero}.  Following previous works \cite{choi2020attentron, casanova2021sc}, we compute SECS using the speaker encoder of the Resemblyzer \cite{Jemine2019Master} package, allowing for comparison with those studies. We also report the Similarity MOS (Sim-MOS) following the works of \cite{jia2018transfer}, \cite{choi2020attentron}, and \cite{casanova2021sc}.
 
Although the experiments involve 3 languages, due to the high cost of the MOS metrics, only two languages were used to compute such metrics: English, which has the largest number of speakers, and Portuguese, which has the smallest number. In addition, following the work of \cite{casanova2021sc} we present such metrics only for speakers unseen during training.

MOS scores were obtained with rigorous crowdsourcing\footnote{https://www.definedcrowd.com/evaluation-of-experience/}. 
For the calculation of MOS and the Sim-MOS in the English language, we use 276 and 200 native English contributors, respectively. For the Portuguese language, we use 90 native Portuguese contributors for both metrics.

During evaluation we use the fifth sentence of the VCTK dataset (i.e, speakerID\_005.txt) as reference audio for the extraction of speaker embeddings, since all test speakers uttered it and because it is a long sentence (20 words). For the LibriTTS and MLS Portuguese, we randomly draw one sample per speaker considering only those with 5 seconds or more, to guarantee a reference with sufficient duration.

For the calculation of MOS, SECS, and Sim-MOS in English, we select 55 sentences randomly from the \textit{test-clean} subset of the LibriTTS dataset, considering only sentences with more than 20 words. For Portuguese we use the translation of these 55 sentences. During inference, we synthesize 5 sentences per speaker in order to ensure coverage of all speakers and a good number of sentences. As ground truth for all test subsets, we randomly select 5 audios for each of the test speakers. For the SECS and Sim-MOS ground truth, we compared such randomly selected 5 audios per speaker with the reference audios used for the extraction of speaker embeddings during synthesis of the test sentences.

Table \ref{tab:results} shows MOS and Sim-MOS with 95\% confidence intervals and SECS for all of our experiments in English for the datasets VCTK and LibriTTS and in Portuguese with the Portuguese sub-set of the dataset MLS. 

\begin{table*}[t]
\centering
\caption{SECS, MOS and Sim-MOS with 95\% confidence intervals for all our experiments.}
\label{tab:results}
\begin{center}
\begin{small}
\begin{sc}
\resizebox{0.95\textwidth}{!}{%
\begin{tabular}{l|c|c|c|c|c|c|c|c|c}
\hline
              & \multicolumn{3}{c|}{\textbf{VCTK}}         & \multicolumn{3}{c|}{\textbf{LibriTTS}}              & \multicolumn{3}{c}{\textbf{MLS-PT}}    \\ \hline
\textbf{Exp.} & \textbf{SECS} & \textbf{MOS}      & \textbf{Sim-MOS}       & \textbf{SECS} & \textbf{MOS} & \textbf{Sim-MOS} & \textbf{SECS} & \textbf{MOS} & \textbf{Sim-MOS} \\ \hline
Ground Truth  &   0.824       &  4.26$\pm$0.04 & 4.19$\pm$0.06 &  0.931        & 4.22$\pm$0.05 &      4.22$\pm$0.06        &   0.9018      & 4.61$\pm$0.05 & 4.41$\pm$0.05 \\ \hline
Attentron ZS  & (0.731)       & (3.86$\pm$0.05) & (3.30 $\pm$0.06) & --            & --  & --           & --            & --  & --           \\ \hline
SC-GlowTTS    & (0.804)        & (3.78$\pm$0.07)   & (3.99$\pm$0.07)   & --            & --  & --           & --            & --  & --           \\ \hline
Exp. 1             & \textbf{0.864}        & 4.21$\pm$0.04 & 4.16$\pm$0.05 & 0.754        & \textbf{4.25$\pm$0.05}&  3.98$\pm$0.07  & --            & --  & --           \\ \hline
Exp. 1 + SCL       & 0.861        & 4.20$\pm$0.05 &  4.13$\pm$0.06 & 0.765        & 4.21$\pm$0.04&  4.05$\pm$0.07 & --            & --  & --           \\ \hline
Exp. 2             & 0.857        & \textbf{4.24$\pm$0.04} &  4.15$\pm$0.06 & 0.762        & 4.22$\pm$0.05& 4.01$\pm$0.07  & 0.740        & 3.96$\pm$0.08  & 3.02$\pm$0.1  \\ \hline
Exp. 2 + SCL       & \textbf{0.864}        & 4.19$\pm$0.05 & \textbf{4.17$\pm$0.06} & 0.773        & 4.23$\pm$0.05&  4.01$\pm$0.07 & 0.745        & 4.09$\pm$0.07  & 2.98$\pm$0.1 \\ \hline
Exp. 3             & 0.851        & 4.21$\pm$0.04 & 4.10$\pm$0.06 & 0.761        & 4.21$\pm$0.04& 4.01$\pm$0.05  & 0.761        & 4.01$\pm$0.08   &  \textbf{3.19$\pm$0.1} \\ \hline
Exp. 3 + SCL       & 0.855        & 4.22$\pm$0.05 & 4.06$\pm$0.06 & 0.778        & 4.17$\pm$0.05 &  3.98$\pm$0.07  & 0.766        & \textbf{4.11$\pm$0.07} & 3.17$\pm$0.1 \\ \hline
Exp. 4 + SCL       & 0.843        & 4.23$\pm$0.05 & 4.10$\pm$0.06  & \textbf{0.856}       & 4.18$\pm$0.05 & \textbf{4.07$\pm$0.07}   & \textbf{0.798}        & 3.97$\pm$0.08 & 3.07$\pm$0.1 \\ \hline
\end{tabular}
}
\end{sc}
\end{small}
\end{center}
\end{table*}

\subsection{VCTK dataset}

For the VCTK dataset, the best similarity results were obtained with experiments 1 (monolingual) and 2 + SCL (bilingual). Both achieved the same SECS and a similar Sim-MOS. According to the Sim-MOS, the use of SCL did not bring any improvements; however, the confidence intervals of all experiments overlap, making this analysis inconclusive. On the other hand, according to SECS, using SCL improved the similarity in 2 out of 3 experiments. Also, for experiment 2, both metrics agree on the positive effect of SCL in similarity. 

Another noteworthy result is that SECS for all of our experiments on the VCTK dataset are higher than the ground truth. This can be explained by characteristics of the VCTK dataset itself which has, for example, significant breathing sounds in most audios. The speaker encoder may not be able to handle these features, hereby lowering the SECS of the ground truth. Overall, in our best experiments with VCTK, the similarity (SECS and Sim-MOS) and quality (MOS) results are similar to the ground truth. Our results in terms of MOS match the ones reported by the VITS article \cite{kim2021conditional}. However, we show that with our modifications, the model manages to maintain good quality and similarity for unseen speakers. Finally, our best experiments achieve superior results in similarity and quality when compared to ~\cite{choi2020attentron, casanova2021sc}; therefore, achieving the SOTA in the VCTK dataset for zero-shot multi-speaker TTS.

\subsection{LibriTTS dataset}
We achieved the best LibriTTS similarity in experiment 4. This result can be explained by the use of more speakers ($\sim1.2$k) than any other experiments ensuring a broader coverage of voice and recording condition diversity.  On the other hand, MOS achieved the best result for the monolingual case. We believe that this was mainly due to the quality of the training datasets. Experiment 1 uses VCTK dataset only, which has higher quality when compared to other datasets added in the other experiments.


\subsection{Portuguese MLS dataset}
For the Portuguese MLS dataset, the highest MOS metric was achieved by experiment 3+SCL, with MOS 4.11$\pm$0.07, although the confidence intervals overlap with the other experiments.  It is interesting to observe that the model trained in Portuguese with a single-speaker dataset of medium quality, manages to reach a good quality in the zero-shot multi-speaker synthesis. Experiment 3 is the best experiment according to Sim-MOS (3.19$\pm$0.10) however, with an overlap with other ones considering the confidence intervals. In this dataset, Sim-MOS and SECS do not agree: based on the SECS metric, the model with higher similarity was obtained in experiment 4+SCL. We believe this is due to the variety in the LibriTTS dataset. The dataset is also composed of audiobooks, therefore tending to have similar recording characteristics and prosody to the MLS dataset. We believe that this difference between SECS and Sim-MOS can be explained by the confidence intervals of Sim-MOS. Finally, Sim-MOS achieved in this dataset is relevant, considering that our model was trained with only one male speaker in the Portuguese language.

Analyzing the metrics by \textbf{gender}, the MOS for experiment 4 considering only male and female speakers are respectively 4.14 $\pm$ 0.11 and 3.79 $\pm$ 0.12. Also, the Sim-MOS for male and female speakers are respectively 3.29 $\pm$ 0.14 and 2.84 $\pm$ 0.14. Therefore, the performance of our model in Portuguese is affected by gender. We believe that this happened because our model was not trained with female Portuguese speakers. Despite that, our model was able to produce female speech in the Portuguese language. The Attentron model achieved a Sim-MOS of 3.30$\pm$0.06 after being trained with approximately 100 speakers in the English language. Considering confidence intervals, our model achieved a similar Sim-MOS even when seeing only one male speaker in the target language. Hence, we believe that our approach can be the solution for the development of zero-shot multi-speaker TTS models in low-resourced languages.



Including \textbf{French} (i.e. experiment 3) appear to have improved both quality and similarity (according to SECS) in Portuguese. The increase in quality can be explained by the fact that the M-AILABS French dataset has better quality than the Portuguese corpus; consequently, as the batch is balanced by language, there is a decrease in the amount of lower quality speech in the batch during model training. Also, increase in similarity can be explained by the fact that TTS-Portuguese is a single speaker dataset and with the batch balancing by language in experiment 2, half of the batch is composed of only one male speaker. When French is added, then only a third of the batch will be composed of the Portuguese speaker voice.

\subsection{Speaker Consistency Loss}

The use of Speaker Consistency Loss (SCL) improved similarity measured by SECS. On the other hand, for the Sim-MOS the confidence intervals between the experiments are inconclusive to assert that the SCL improves similarity. Nevertheless, we believe that SCL can help the generalization in recording characteristics not seen in training. For example, in experiment 1, the model did not see the recording characteristics of the LibriTTS dataset in training but during testing on this dataset, both the SECS and Sim-MOS metrics showed an improvement in similarity thanks to SCL. On the other hand, it seems that using SCL slightly decreases the quality of generated audio. We believe this is because with the use of SCL, our model learns to generate recording characteristics present in the reference audio, producing more distortion and noise. However, it should be noted that in our tests with high-quality reference samples, the model is able to generate high-quality speech.

\section{Zero-Shot Voice Conversion}

\begin{table*}[t]
\centering
\caption{MOS and Sim-MOS with 95\% confidence intervals for the zero-shot voice conversion experiments.}
\label{tb:vc}
\begin{center}
\begin{small}
\begin{sc}
\resizebox{\textwidth}{!}{%
\begin{tabular}{c|c|c|c|c|c|c|c|c|c|c}
\hline

\multirow{2}{*}{\textbf{\begin{tabular}[c]{@{}c@{}}Ref/Tar\end{tabular}}} & \multicolumn{2}{c|}{\textbf{M-M}}    & \multicolumn{2}{c|}{\textbf{M-F}}    & \multicolumn{2}{c|}{\textbf{F-F}}    & \multicolumn{2}{c|}{\textbf{F-M}} & \multicolumn{2}{c}{\textbf{All}}     \\ \cline{2-11} 
                & MOS    & Sim-MOS  & MOS       & Sim-MOS  & MOS       & Sim-MOS  & MOS        & Sim-MOS     & MOS        & Sim-MOS        \\ \hline
        
\textbf{en-en}  & 4.22$\pm$0.10 & 4.15$\pm$0.12 & 4.14$\pm$0.09 & 4.11$\pm$0.12 & 4.16$\pm$0.12  & 3.96$\pm$0.15 & 4.26$\pm$0.09  & 4.05$\pm$0.11 & 4.20$\pm$0.05 &  4.07$\pm$0.06 \\ \hline
\textbf{pt-pt}  & 3.84 $\pm$ 0.18 & 3.80 $\pm$ 0.15 & 3.46 $\pm$ 0.10 & 3.12 $\pm$ 0.17   & 3.66 $\pm$ 0.2 &  3.35 $\pm$ 0.19 & 3.67 $\pm$ 0.16 &  3.54 $\pm$ 0.16 & 3.64 $\pm$ 0.09 & 3.43 $\pm$ 0.09 \\ \hline
\textbf{en-pt}  & 4.17$\pm$0.09 &  3.68 $\pm$ 0.10 &  4.24$\pm$0.08    & 3.54 $\pm$ 0.11  &  4.14$\pm$0.09     & 3.58 $\pm$ 0.12 & 4.12$\pm$0.10     & 3.58 $\pm$ 0.11 &  4.17$\pm$0.04 & 3.59 $\pm$ 0.05 \\ \hline
\textbf{pt-en}  &   3.62 $\pm$ 0.16   & 3.8 $\pm$ 0.10 &    2.95 $\pm$ 0.2  &3.67 $\pm$ 0.11  & 3.51 $\pm$ 0.18  &  3.63 $\pm$ 0.11 & 3.47 $\pm$ 0.18 & 3.57 $\pm$ 0.11 &   3.40 $\pm$ 0.09 & 3.67 $\pm$ 0.05 \\ \hline
\end{tabular}
}
\end{sc}
\end{small}
\end{center}
\end{table*}

As in the SC-GlowTTS \cite{casanova2021sc} model, we do not provide any information about the speaker’s identity to the encoder, so the distribution predicted by the encoder is forced to be speaker independent. Therefore, YourTTS can convert voices using the model’s Posterior Encoder, decoder and the HiFi-GAN Generator. Since we conditioned YourTTS with external speaker embeddings, it enables our model to mimic the voice of unseen speakers in a zero-shot voice conversion setting. 

In \cite{wang2021noisevc}, the authors reported the MOS and Sim-MOS metrics for the AutoVC  \cite{qian2019autovc} and NoiseVC \cite{wang2021noisevc} models for 10 VCTK speakers not seen during training. To compare our results, we selected 8 speakers (4M/4F) from the VCTK test subset. Although \cite{wang2021noisevc} uses 10 speakers, due to gender balance, we were forced to use only 8 speakers.

Furthermore, to analyze the generalization of the model for the Portuguese language, and to verify the result achieved by our model in a language where the model was trained with only one speaker, we used the 8 speakers (4M/4F) from the test subset of the MLS Portuguese dataset. Therefore, in both languages we use speakers not seen in the training. Following \cite{qian2019autovc} for a deeper analysis, we compared the transfer between male, female and mixed gender speakers individually. During the analysis, for each speaker, we generate a transfer in the voice of each of the other speakers, choosing the reference samples randomly, considering only samples longer than 3 seconds.  In addition, we analyzed voice transfer between English and Portuguese speakers.  We calculate the MOS and the Sim-MOS as described in Section \ref{sec:results}. However, for the calculation of the sim-MOS when transferring between English and Portuguese (pt-en and en-pt), as the reference samples are in one language and the transfer is done in another language, we used evaluators from both languages (58 and 40, respectively, for English and  Portuguese).

Table \ref{tb:vc} presents the MOS and Sim-MOS for these experiments. Samples of the zero-shot voice conversion are present in the demo page\footnote{https://edresson.github.io/YourTTS/}.


        

\subsection{Intra-lingual results}
For zero-shot voice conversion from one English-speaker to another English-speaker (en-en) our model achieved a MOS of 4.20$\pm$0.05 and a Sim-MOS of 4.07$\pm$0.06. 
For comparison in \cite{wang2021noisevc} the  authors reported the MOS and Sim-MOS results for the AutoVC \cite{qian2019autovc} and NoiseVC \cite{wang2021noisevc} models. For 10 VCTK speakers not seen during training, the AutoVC model achieved a MOS of $3.54\pm 1.08$\footnote{The authors presented the results in a graph without the actual figures, so the MOS scores reported here are approximations calculated considering the length in pixels of those graphs.} and a Sim-MOS of $1.91\pm 1.34$. On the other hand, the NoiseVC model achieved a MOS of $3.38\pm 1.35$ and a Sim-MOS of $3.05\pm 1.25$. Therefore, our model achieved results comparable to the SOTA in zero-shot voice conversion in the VCTK dataset. Alhtough the model was trained with more data and speakers, the similarity results of the VCTK dataset in Section \ref{sec:results} indicate that the model trained with only the VCTK dataset (experiment 1) presents a better similarity than the model explored in this Section (experiment 4). Therefore, we believe that YourTTS can achieve a result very similar or even superior in zero-shot voice conversion when being trained and evaluated using only the VCTK dataset.

For zero-shot voice conversion from one Portuguese speaker to another Portuguese speaker our model achieved a MOS of 3.64 $\pm$ 0.09 and a Sim-MOS of 3.43 $\pm$ 0.09. 
We note that our model performs significantly worse in voice transfer similarity between female speakers (3.35 $\pm$ 0.19) compared to transfers between male speakers (3.80 $\pm$ 0.15). This can be explained by the lack of female speakers for the Portuguese language during the training of our model. Again, it is remarkable that our model manages to approximate female voices in Portuguese without ever having seen a female voice in that language.

\subsection{Cross-lingual results}

Apparently, the transfer between English and Portuguese speakers works as well as the transfer between Portuguese speakers. However, for the transfer of a Portuguese speaker to an English speaker (pt-en) the MOS scores drop in quality. This was especially due to the low quality of voice conversion from Portuguese male speakers to English female speakers. In general, as discussed above, due to the lack of female speakers in the training of the model, the transfer to female speakers achieves poor results. In this case, the challenge is even greater as it is necessary to convert audios from a male speaker in Portuguese to the voice of a English female speaker.

In English, during conversions, the speaker's gender did not significantly influence the model's performance. However, for transfers involving Portuguese, the absence of female voices in the training of the model hindered generalization.





\section{Speaker Adaptation}

The different recording conditions are a challenge for the generalization of the zero-shot multi-speaker TTS models. Speakers who have a voice that differs greatly from those seen in training also become a challenge \cite{tan2021survey}. Nevertheless, to show the potential of our model for adaptation to new speakers/recording conditions, we selected samples from 20 to 61 seconds of speech for 2 Portuguese and 2 English speakers (1M/1F) in the Common Voice \cite{ardila2020common} dataset. Using these 4 speakers, we perform fine-tuning on the checkpoint from experiment 4 with Speaker Consistency Loss individually for each speaker. 

During fine-tuning, to ensure that multilingual synthesis is not impaired, we use all the datasets used in experiment 4. However, we use Weighted random sampling \cite{paszke2017automatic} to guarantee that samples from adapted speakers appear in a quarter of the batch. The model is trained that way for 1500 steps. For evaluation, we use the same approach described in Section \ref{sec:results}. 

\begin{table*}[t]
\centering
\caption{SECS, MOS and Sim-MOS with 95\% confidence intervals for the speaker adaptation experiments.}
\label{tb:spk_adpt}
\begin{center}
\begin{small}
\begin{sc}
\begin{tabular}{c|c|c|c|c|c|c}
\hline
  & \textbf{Sex}    & \textbf{ Dur. (Sam.)} & \textbf{Mode} & \textbf{SECS} & \textbf{MOS} & \textbf{Sim-MOS} \\ \hline
\multirow{6}{*}{EN} & \multirow{3}{*}{M} & \multirow{3}{*}{61s (15)}        & GT            & 0.875         & 4.17$\pm$0.09 &      \textbf{4.08$\pm$0.13}        \\ \cline{4-7} 
  &  &             & ZS                                                     & 0.851         & 4.11$\pm$0.07 &   4.04$\pm$0.09           \\ \cline{4-7} 
  &  &             & FT                                                     & \textbf{0.880}         & 4.17$\pm$0.07 &   \textbf{4.08$\pm$0.09}            \\ \clineB{2-7}{2.5}
  & \multirow{3}{*}{F} & \multirow{3}{*}{44s (11)}        & GT            & 0.894         & 4.25$\pm$0.11 &   \textbf{4.17$\pm$0.13}           \\ \cline{4-7} 
  &  &             & ZS                                                     & 0.814         & 4.12$\pm$0.08 &   4.11$\pm$0.08  \\ \cline{4-7}
  &  &             & FT                                                     &\textbf{ 0.896}         & 4.10$\pm$0.08 &  \textbf{4.17$\pm$0.08}            \\ \hline \clineB{1-7}{1.5} 
\multirow{6}{*}{PT} & \multirow{3}{*}{M} & \multirow{3}{*}{31s (7)}         & GT            & 0.880         & 4.76$\pm$0.12  & \textbf{4.31$\pm$0.14}             \\ \cline{4-7}
  &  &             & ZS                                                     & 0.817         & 4.03$\pm$0.11  & 3.35$\pm$0.12             \\ \cline{4-7} 
  &  &             & FT                                                      & \textbf{0.915}         & 3.74$\pm$0.12 & 4.19$\pm$0.07 \\ \clineB{2-7}{2.5} 
  & \multirow{3}{*}{F} & \multirow{3}{*}{20s  (5)}        & GT            & 0.873         & 4.62$\pm$0.19  & \textbf{4.65$\pm$0.14}             \\ \cline{4-7} 
  &  &             & ZS                                                     & 0.743         & 3.59$\pm$0.13  &  2.77$\pm$0.15         \\ \cline{4-7} 
  &  &             & FT                                                     & \textbf{0.930}         & 3.48$\pm$0.13 & 4.43$\pm$0.06             \\ \hline
\end{tabular}
\end{sc}
\end{small}
\end{center}

\end{table*}

Table \ref{tb:spk_adpt} shows the gender, total duration in seconds and number of samples used during the training for each speaker, and the metrics SECS, MOS and Sim-MOS for the ground truth (GT), zero-shot multi-speaker TTS mode (ZS), and the fine-tuning (FT) with speaker samples.



In general, our model's fine-tuning with less than 1 minute of speech from speakers who have recording characteristics not seen during training achieved very promising results, significantly improving similarity in all experiments.

In English, the results of our model in zero-shot multi-speaker TTS mode are already good and after fine-tuning both male and female speakers achieved Sim-MOS comparable to the ground truth. The fine-tuned model achieves greater SECS than the ground truth, which was already observed in previous experiments. We believe that this phenomenon can be explained by the model learning to copy the recording characteristics and reference sample's distortions, giving an advantage over other real speaker samples.

In Portuguese, compared to zero-shot, fine-tuning seems to trade a bit of naturalness for a much better similarity. For the male speaker, the Sim-MOS increased from 3.35$\pm$0.12 to 4.19$\pm$0.07 after fine-tuning with just 31 seconds of speech for that speaker. For the female speaker, the similarity improvement was even more impressive, going from 2.77$\pm$0.15 in zero-shot mode to 4.43$\pm$0.06 after the fine-tuning with just 20 seconds of speech from that speaker.

Although our model manages to achieve high similarity using only seconds of the target speaker's speech, Table \ref{tb:spk_adpt} seems to presents a direct relationship between the amount of speech used and the naturalness of speech (MOS). With approximately 1 minute of speech in the speaker's voice our model can copy the speaker's speech characteristics, even increasing the naturalness compared to zero-shot mode. On the other hand, using 44 seconds or less of speech reduces the quality/naturalness of the generated speech when compared to the zero-shot or ground truth model. Therefore, although our model shows good results in copying the speaker's speech characteristics using only 20 seconds of speech, more than 45 seconds of speech are more adequate to allow higher quality. Finally, we also noticed that voice conversion improves significantly after fine-tuning the model, mainly in Portuguese and French where few speakers are used in training.

\section{Conclusions, limitations and future work} \label{sec:conc}

In this work, we presented YourTTS, which achieved SOTA results in zero-shot multi-speaker TTS and zero-shot voice conversion in the VCTK dataset. Furthermore, we show that our model can achieve promising results in a target language using only a single speaker dataset. Additionally, we show that for speakers who have both a voice and recording conditions that differ greatly from those seen in training, our model can be adjusted to a new voice using less than 1 minute of speech. 

{However, our model exhibits some limitations. For the TTS experiments in all languages, our model presents instability in the stochastic duration predictor which, for some speakers and sentences, generates unnatural durations. 
We also note that mispronunciations occur for some words, especially in Portuguese. Unlike \cite{casanova2020ttsportuguese, casanova2020end, kim2021conditional}, we do not use phonetic transcriptions, making our model more prone to such problems.  
For Portuguese voice conversion, the speaker's gender significantly influences the model's performance,  due to the absence of female voices in training. 
For Speaker Adaptation, although our model shows good results in copying the speaker's speech characteristics using only 20 seconds of speech, more than 45 seconds of speech are more adequate to allow higher quality.}

In future work, we intend to seek improvements to the duration predictor of the YourTTS model as well as training in more languages. Furthermore, we intend to explore the application of this model for data augmentation in the training of automatic speech recognition models in low-resource settings.

\section{Acknowledgements}

This study was financed in part by the Coordena\c{c}\~{a}o de Aperfei\c{c}oamento de Pessoal de N\'{i}vel Superior -- Brasil (CAPES) -- Finance Code 001, as well as CNPq (National Council of Technological and Scientific Development) grants 304266/2020-5. In addition, this research was financed in part by  Artificial Intelligence Excellence Center (CEIA)\footnote{\url{http://centrodeia.org}} via projects funded by the Department of Higher Education of the Ministry of Education (SESU/MEC) and Cyberlabs Group\footnote{\url{https://cyberlabs.ai}}. Also, we would like to thank the Defined.ai\footnote{\url{https://www.defined.ai}} for making industrial-level MOS testing so easily available. Finally, we would like to thank all contributors to the Coqui TTS repository\footnote{\url{https://github.com/coqui-ai/TTS}}, this work was only possible thanks to the commitment of all.

\bibliographystyle{RefStyle}

\bibliography{references}

\clearpage
\appendix

\section{Erratum}

In Section \ref{sec:TTSModel} of this paper, we have defined the Speaker Consistency Loss (SCL) function. In addition, we have used this loss function on 4 fine-tuning experiments in Sections \ref{sec:method} and \ref{sec:results} (EXP. 1 + SCL, EXP. 2 + SCL, EXP. 3 + SCL,  and EXP. 4 + SCL). However, due to an implementation mistake, the gradient of this loss function was not propagated for the model during the training. It means that the fine-tuning experiments that  used this loss are equivalent to training the model for more steps without the Speaker Consistency Loss. This bug was discovered by Tomáš Nekvinda\footnote{https://github.com/Tomiinek}  and reported on issue number  2348  of the Coqui TTS repository\footnote{https://github.com/coqui-ai/TTS/issues/2348}.  This bug was fixed on the pull request number 2364 on the Coqui TTS repository\footnote{https://github.com/coqui-ai/TTS/pull/2364}. Currently,  it is fixed for Coqui TTS version v0.12.0 or higher.  We would like to thank Tomáš Nekvinda for finding the bug and reporting it.

\end{document}